\documentclass[12pt]{article}

\usepackage{epsfig}

\begin{document}

\title{\bf \Large Energy of a Regular Black Hole}

\author{M. Sharif \thanks{Present Address: Department of Mathematical Sciences,
University of Aberdeen, Kings College, Aberdeen AB24 3UE Scotland,
UK. $<$msharif@maths.abdn.ac.uk$>$}
\\ Department of Mathematics, University of the Punjab,
\\ Quaid-e-Azam Campus Lahore-54590, Pakistan.\\
$<$hasharif@yahoo.com$>$}

\date{}

\maketitle

\begin{abstract}
We use Einstein, Landau-Lifshitz, Papapetrou and Weinberg
energy-momentum complexes to evaluate energy distribution of a
regular black hole. It is shown that for a regular black hole,
these energy-momentum complexes give the same energy distribution.
This supports Cooperstock hypothesis and also Aguirregabbiria et
al. conclusions. Further, we evaluate energy distribution using
M$\ddot{o}$ller's prescription. This does not exactly coincide
with ELLPW energy expression but, at large distances, they become
same.
\end{abstract}


\maketitle

\section{Introduction}

The definition of energy-momentum has always been a focus of many
investigations in General Relativity (GR). This, together with
conservation laws, has a crucial role in any physical theory.
However, there is still no accepted definition of energy-momentum,
and generally speaking, of conserved quantities associated with
the gravitational field. The main difficulty is with the
expression which defines the gravitational field energy part.
Einstein used principle of equivalence and conservation laws of
energy-momentum to formulate the covariant field equations. He
formulated the energy-momentum conservation law in the form
\begin{equation}
\frac{\partial}{\partial x^b}(\sqrt{-g}(T^b_a+t^b_a))=0,\quad
(a,b=0,1,2,3),
\end{equation}
where $T^b_a$ is the stress energy density of matter. He
identified $t^b_a$ as representing the stress energy density of
gravitation. He noted that $t^b_a$ was not a tensor, but concluded
that the above equations hold good in all coordinate systems since
they were directly obtained from the principle of GR. The choice
of a non-tensorial quantity to describe the gravitational field
energy immediately attracted some criticism.

The problems associated with Einstein's pseudo-tensor resulted in
many alternative definitions of energy, momentum and angular
momentum being proposed for a general relativistic system. These
include Landau-Lifshitz$^1$, Tolman$^2$, Papapetrou$^3$,
Bergmann$^4$, Weinberg$^5$ who had suggested different expressions
for the energy-momentum distribution. The main problem with these
definitions is that they are coordinate dependent. One can have
meaningful results only when calculations are performed in
Cartesian coordinates. This restriction of coordinate dependent
motivated some other physicists like M$\ddot{o}$ller$^{6,7}$,
Komar$^8$ and Penrose$^9$ who constructed coordinate independent
definitions of energy-momentum complex.

M$\ddot{o}$ller claimed that his expression gives the same values
for the total energy and momentum as the Einstein's
energy-momentum complex for a closed system. However,
M$\ddot{o}$ller's energy-momentum complex was subjected to some
criticism$^{7-10}$. Komar's prescription, though not restricted to
the use of Cartesian coordinates, is not applicable to non-static
spacetimes. Penrose$^9$ pointed out that quasi-local masses are
conceptually very important. However, different definitions of
quasi-local masses do not give agreed results for the
Reissner-Nordstrom and Kerr metrics and that the Penrose
definition could not succeed to deal with the Kerr. These
inadequacies of quasi-local definitions have been discussed in a
series of papers$^{11-13}$. Thus each of these energy-momentum
complex has its own drawback. As a result these ideas of the
energy-momentum complex were severally criticized.

Virbhadra$^{13,14}$ revived the interest in this approach by
showing that different energy-momentum complexes can give the same
energy-momentum. Since then lot of work on evaluating the
energy-momentum distributions of different spacetimes have been
carried out by different authors$^{15-18}$. In a recent paper,
Virbhadhra$^{13}$ used the energy-momentum complexes of Einstein,
Landau-Lifshitz, Papapetrou and Weinberg (ELLPW) to investigate
whether or not they can give the same energy distribution for the
most general non-static spherically symmetric metric. It was a
great surprise that contrary to previous results of many
asymptotically flat spacetimes and asymptotically non-flat
spacetimes, he found that these definitions disagree. He observed
that Einstein's energy-momentum complex provides a consistent
result for the Schwarzschild metric whether one calculates in
Kerr-Schild Cartesian coordinates or Schwarzschild Cartesian
coordinates. The prescriptions of Landau-Lifshitz, Papapetrou and
Weinberg furnish the same result as in the Einstein prescription
if the calculations are carried out in Schwarzschild Cartesian
coordinates. Thus the prescriptions of Landau-Lifshitz, Papapetrou
and Weinberg do not give a consistent result. On the basis of
these and some other facts$^{11,12}$, Virbhadra concluded that the
Einstein method seems to be the best among all known (including
quasi-local mass definitions) for energy distribution in a
spacetime. Recently, Lessner$^{19}$ pointed out that the
M$\ddot{o}$ller's energy-momentum prescription is a powerful
concept of energy and momentum in GR.

It has been shown recently$^{13}$ that ELLPW energy-momentum
complexes coincide for any Kerr-Schild class metric when one uses
Kerr-Schild Cartesian coordinates. In this paper we use ELLPW and
M$\ddot{o}$ller energy-momentum complexes to obtain the energy
distribution of a regular black hole which is represented by a
Bardeen's model$^{20, 21}$. It is shown that ELLPW energy-momentum
complexes give the same and acceptable results for a given
space-time. Our results agree with Virbhadra's conclusion that the
Einstein's energy-momentum complex is still the best tool for
obtaining energy distribution in a given spacetime. This also
supports Cooperstock's hypothesis (that energy and momentum in a
curved space-time are confined to the the regions of non-vanishing
energy-momentum of matter and the non-gravitational field).

The paper has been organised as follows. In the next section, we
shall describe the regular black holes. In Secs. 3 and 4, we
evaluate energy distribution using ELLPW and M$\ddot{o}$ller's
prescriptions respectively. Finally, we shall discuss the results.

\section{Regular Black Holes}

In 1968, Bardeen$^{20,21}$ constructed a well-known model called
Bardeen's model. This model represents a regular black hole
obeying the weak energy condition, and it was powerful in shaping
the direction of research on the existence or avoidance of
singularities. The model uses the Reissner-Nordstr$\ddot{o}$m
spacetime as inspiration. The metric expressed in standard
spherical coordinates $(t,r,\theta,\phi)$ is given by the line
element of the form
\begin{equation}
ds^2=f(r)dt^2-f^{-1}(r)dr^2-r^2(d\theta^2+\sin^2\theta d\phi^2),
\end{equation}
where Bardeen replaced the usual Reissner-Nordstr$\ddot{o}$m
function
\begin{equation}
f(r)=1-\frac{2m}{r}+\frac{e^2}{r^2}
\end{equation}
by
\begin{equation}
f(r)=1-\frac{2mr^2}{(r^2+e^2)^{3/2}}.
\end{equation}
When $e^2<(16/27)m^2$ in Bardeen's model, there is an event
horizon. There are values $r_\pm$ of $r$ such that the region
$r_-<r<r_+$ contains trapped surfaces. The spacetime obeys the
null convergence, yet it contains no physical singularities. It is
to be noticed that if we take charge $e=0$, the above metric
reduces to the Schwarzschild metric.

\section{Energy of the Regular Black Hole}

In this section we shall use ELLPW energy-momentum complexes to
evaluate the energy distribution of the regular black hole. To
this end, we shall follow the procedure developed by
Virbhadra$^{13}$. The basic requirement of the procedure is to
bring the metric in the form of Kerr-Schild class and then
transform the resulting metric in Kerr-Schild Cartesian
coordinates.

The metrics of the Kerr-Schild class are written in the following
form
\begin{equation}
g_{ab}=\eta_{ab}-Hl_al_b,
\end{equation}
where $\eta_{ab}$ is the Minkowski spacetime, $H$ is the scalar
field and $l_a$ is a null, geodesic and shear free vector field in
the Minkowski metric. These can be expressed as
\begin{equation}
\eta^{ab}l_al_b=0, \quad \eta^{ab}l_{c,a}l_b=0,\quad
(l_{a,b}+l_{b,a})l^a_{,c}\eta^{bc}-(l^a_{,a})^2=0,
\end{equation}
where $a,b,c=0,1,2,3$. It is to be noticed that, for the
Kerr-Schild class metric, the vector field $l_a$ remains null,
geodesic and shear free with the metric $g_{ab}$. Thus it follows
from the above equation that
\begin{equation}
g^{ab}l_al_b=0, \quad g^{ab}l_{c;a}l_b=0,\quad
(l_{a;b}+l_{b;a})l^a_{;c}g^{bc}-(l^a_{;a})^2=0.
\end{equation}

Now we bring the metric given by Eq.(2) in Kerr-Schild class by
using the following coordinate transformation
\begin{equation}
u=t+\int{f^{-1}}(r)dr
\end{equation}
which implies that
\begin{equation}
ds^2=f(r)du^2-2dudr-r^2(d\theta^2+\sin^2\theta d\phi^2).
\end{equation}
This metric turns out to be static case of the Kerr-Schild class
as given by Aguirregabiria et al.$^{13}$.

In order to have meaningful results in the prescriptions of ELLPW,
it is necessary to transform the metric in Kerr-Schild Cartesian
coordinates. Let us now transform the metric in Kerr-Schild
Cartesian coordinates by using
\begin{equation}
T=u-r, \quad x=r\sin\theta\cos\phi,\quad y=r\sin\theta\sin\phi,
\quad z=r\cos\theta.
\end{equation}
The corresponding metric in these coordinates will become
\begin{equation}
ds^2=dT^2dx^2-dy^2-dz^2-(1-f(r))[dT+\frac{1}{r}(xdx+ydy+zdz)]^2.
\end{equation}
This is the Kerr-Schild class metric with $H=1-f$ and $l_a=(1,
\frac{x}{r},\frac{y}{r}\frac{z}{r})$. We use the procedure of
Aguirregabiria et al.$^{13}$ to calculate energy distribution of
the regular black hole in the ELLPW prescriptions. It turns out
that we get the same energy in these prescriptions which is given
as
\begin{equation}
E_{ELLPW}=\frac{r}{2}(1-f).
\end{equation}
When we replace the value of $f$ from Eq.(4), it follows that the
energy distribution of the regular black hole is
\begin{equation}
E_{ELLPW}=\frac{mr^3}{(r^2+e^2)^{3/2}}
\end{equation}
which can be written as follows
\begin{eqnarray}
E_{ELLPW}=m(1-\frac{3e^2}{2r^2}+\frac{15e^4}{8r^4}+O(\frac{1}{r^6})).
\end{eqnarray}
If we take the charge $e=0$ or at large distances, it reduces to
the energy of the Schwarzscild metric given by
\begin{equation}
E_{ELLPW}=m
\end{equation}

\section{Energy Distribution in M$\ddot{o}$ller's Prescription}

Now we shall use M$\ddot{o}$ller's energy-momentum complex to
evaluate energy of the regular black hole. This is the beauty of
M$\ddot{o}$ller's method that it is independent of coordinates and
consequently we can perform computations in spherical polar
coordinates. The energy-momentum complex of
M$\ddot{o}$ller$^{6,7}$ is given by
\begin{equation}
M_a^b=\frac{1}{8\pi}H^{bc}_{a,c}~,
\end{equation}
where
\begin{equation}
H_a^{bc}=\sqrt{-g}(g_{ad,e}-g_{ae,d})g^{be}g^{cd},\quad
(a,b,c,d=0,1,2,3).
\end{equation}
The energy-momentum complex satisfies the local conservation laws
\begin{equation}
\frac{\partial M_a^b}{\partial x^b}=0.
\end{equation}
The locally conserved energy-momentum complex $M_a^b$ contains
contributions from the matter, non-gravitational and gravitational
fields. $M_0^0$ and $M_a^0$ are the energy and momentum (energy
current) density components respectively. The energy and momentum
components are given as
\begin{equation}
P_\alpha=\int\int\int{M_\alpha^0dx^1dx^2dx^3},\quad
(\alpha=0,1,2,3),
\end{equation}
where $P_0$ is the energy and $P_i$ represent the momentum
components. Using Gauss's theorem, the energy expression E can be
written as
\begin{equation}
E=\frac{1}{8\pi}\int\int H_0^{0j}n_jdS,\quad (i,j=1,2,3),
\end{equation}
where $n_j$ is the outward unit normal vector over an infintesimal
surface element $dS$.

The only required component of $H_a^{bc}$, to evaluate energy of
the Bardeen's model, is given by
\begin{equation}
H_0^{01}=\frac{2mr^3(r^2-2e^2)\sin\theta}{(r^2+e^2)^{5/2}}.
\end{equation}
Substituting this value of $H_0^{01}$ in Eq.(19), we have the
following energy distribution in M$\ddot{o}$ller's prescription
\begin{eqnarray}
E_M=\frac{mr^3(r^2-2e^2)}{(r^2+e^2)^{5/2}}
\end{eqnarray}
which can be written as
\begin{eqnarray}
E_M=m(1-\frac{9e^2}{2r^2}+\frac{75e^4}{8r^4}+O(\frac{1}{r^6})).
\end{eqnarray}
We see that the energy expression for ELLPW and M$\ddot{o}$ller's
prescriptions coincide at large distances. They are exactly the
same for the Schwarzschild metric.

\section{Discussion}

There are two types of energy-momentum complexes in the
literature. The first type depends on the coordinates and the
other type is independent of coordinate. However, it has been
shown by many authors that the first type give more meaning
results. The debate on the localization of energy-momentum is also
an interesting and a controversial problem. According to Misner et
al$^{22}$, energy can only be localized for spherical systems.
However, Cooperstock and Sarracino$^{23}$ suggested that if energy
can be localized in spherical systems then it can be localized in
any spacetimes. The energy-momentum complexes are non-tensorial
under general coordinate transformations and hence are restricted
to Cartesian coordinates only. In their recent work Virbhadra and
his collaborators$^{13-18}$ have shown that different
energy-momentum complexes can provide meaningful results.

In this paper, we have evaluated energy of the regular black hole
using prescriptions of ELLPW. It is worth noting that the energy
turns out to be same in the prescriptions of Einstein,
Landau-Lifshitz, Papapetrou and Weinberg. It is clear that the
definitions of ELPPW support the Cooperstock hypothesis for the
regular black hole. We have also calculated this quantity using
M$\ddot{o}$ller energy-momentum complex. This is not exactly the
same as evaluated by using ELLPW prescriptions. However, it can be
seen from Eqs.(14) and (23) that, at large distances, these give
the same result and reduces to the energy of Schwarzscild
spacetime.

We plot the energy distributions of ELLPW ($E_{ELLLPW}/m$) and
M$\ddot{o}$ller ($E_M/m$) in the figures 1 and 2 respectively.

\begin{figure}
\begin{center}
\epsfig{file=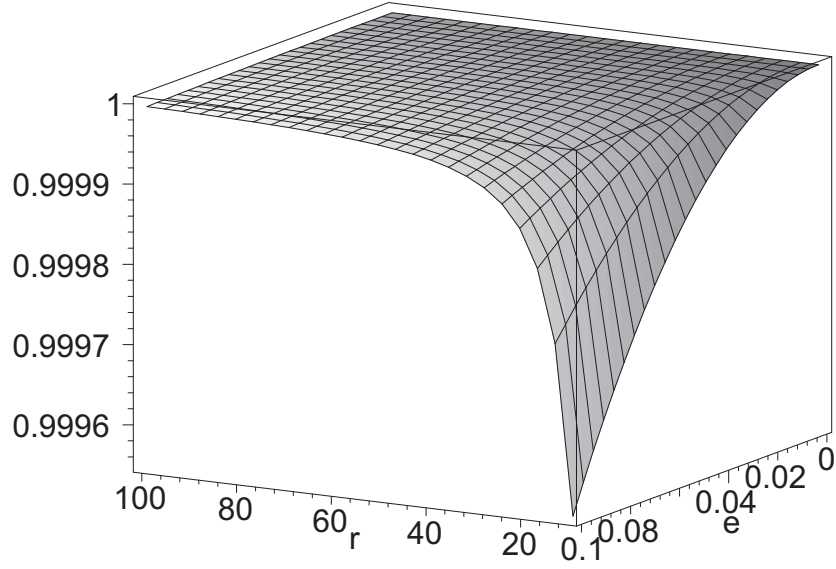,width=0.8\linewidth}
\end{center}
\caption{$E_{ELLPW}/m$ on z-axis are plotted against r on x-axis
and e on y-axis.}
\end{figure}
\bigskip
\begin{figure}
\begin{center}
\epsfig{file=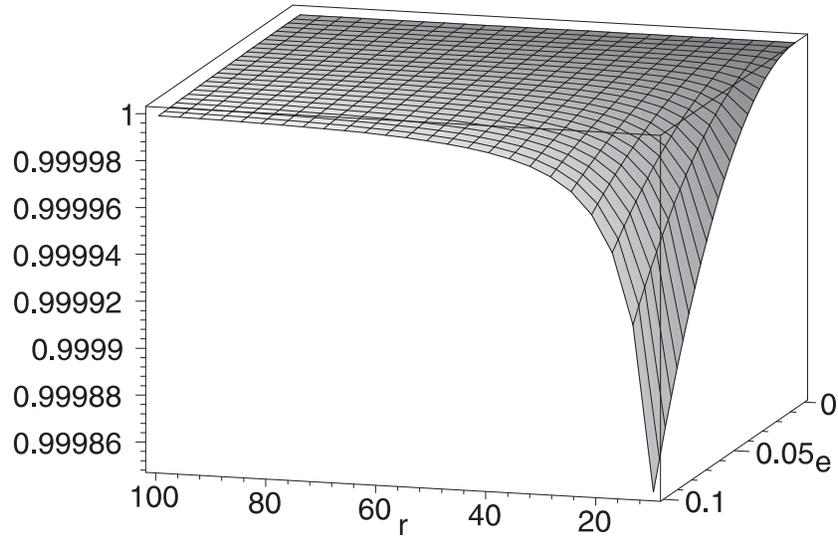,width=0.8\linewidth}
\end{center}
\caption{$E_M/m$ on z-axis are plotted against r on x-axis and e
on y-axis.}
\end{figure}

\newpage

\begin{description}
\item  {\bf Acknowledgment}
\end{description}

I would like to thank Ministry of Science and Technology (MOST),
Pakistan for providing postdoctoral fellowship at University of
Aberdeen, UK.

\vspace{2cm}

{\bf \large References}

\begin{description}

\item{1.} L.D. Landau and E.M. Lifshitz, {\it The Classical Theory
of Fields} (Addison-Wesley Press, Reading, MA, 1962)2nd ed.

\item{2.} R.C. Tolman, {\it Relativity, Thermodynamics and
Cosmology} (Oxford Univ. Press, 1934)227.

\item{3.} A. Papapetrou, {\it Proc. R. Irish. Acad.} {\bf A52}, 11
(1948).

\item{4.} P.G. Bergmann and R. Thompson, {\it Phys. Rev.} {\bf
89}, 400 (1953).

\item{5.} S. Weinberg, {\it Gravitation and Cosmology} (Wiley, New
York, 1972).

\item{6.} C. M$\ddot{o}$ller, {\it Ann. Phys. (NY)} {\bf 4}, 347
(1958).

\item{7.} C. M$\ddot{o}$ller, {\it Ann. Phys. (NY)} {\bf 12}, 118
(1961).

\item{8.} A. Komar, {\it Phys. Rev.} {\bf 113}, 934 (1959).

\item{9.} R. Penrose, {\it Proc. Roy. Soc. London} {\bf A381}, 53
(1982).

\item{10.} D. Kovacs, {\it Gen. Relatv. and Grav.} {\bf 17}, 927
(1985); J. Novotny, {\it Gen. Relatv. and Grav.} {\bf 19}, 1043
(1987).

\item{11.} G. Bergqvist, {\it Class. Quantum Grav.} {\bf 9}, 1753
(1992).

\item{12.} D.H. Bernstein and K.P. Tod, {\it Phys. Rev.} {\bf
D49}, 2808 (1994).

\item{13.} K.S. Virbhadra, {\it Phys. Rev.} {\bf D60}, 104041
(1999).

\item{14.} K.S. Virbhadra, {\it Phys. Rev.} {\bf D41}, 1081
(1990); {\bf D42}, 1066 (1990); {\bf D42}, 2919 (1990) and
references therein.

\item{15.} S.S. Xulu, {\it Int. J. Mod. Phys.} {\bf A15}, 2979
(2000); {\it Mod. Phys. Lett.} {\bf A15}, 1511 (2000) and
references therein.

\item{16.} I.C. Yang and I. Radinschi, {\it Mod. Phys. Lett.} {\bf
A17}, 1159 (2002).

\item{17.} M. Sharif, {\it Int. J. of Mod. Phys.} {\bf A17}, 1175
(2002).

\item{18.} M. Sharif, {\it Int. J. of Mod. Phys.} {\bf A18},
(2003); Erratum {\bf A}, (2003); {\bf gr-qc/0310018}.

\item{19.} G. Lessner, {\it Gen. Relativ. Grav.} {\bf 28}, 527
(1996).

\item{20.} J. Bardeen, {\it Proc. GR5} (Tiflis, USSR, 1968).

\item{21.} A. Borde, {\it Phys. Rev.} {\bf D 50}, 3692 (1994);
{\it Phys. Rev.} {\bf D 55}, 7615 (1997).

\item{22.} C.W. Misner, K.S. Thorne and J.A. Wheeler, {\it
Gravitation} (W.H. Freeman, New York, 1973)603.

\item{23.} F.I. Cooperstock and R.S. Sarracino, {\it J. Phys. A.:
Math. Gen.} {\bf 11}, 877 (1978).

\end{description}

\end{document}